\documentclass{aa}

\makeatletter
\providecommand{\aa@subabstractivname}{Results}
\makeatother

\usepackage{txfonts}

\usepackage[english]{babel}

\usepackage{amsmath}
\usepackage{multirow}
\usepackage{threeparttable}
\usepackage{cancel}
\usepackage{orcidlink}

\usepackage[dvipsnames]{xcolor}
\usepackage{hyperref}
\hypersetup{
    pdftitle={The double quasar},
    pdfauthor={Minghui Xu},
    colorlinks=true,       
    linkcolor=MidnightBlue,
    filecolor=magenta,      
    urlcolor=violet,
    citecolor=teal,        
    pdfborder={0 0 0}
}
\title{Registering Gaia positions on VLBI images: constraining opacity-driven core shift in B\,1038+528\,/\,B\,1038+529}
\titlerunning{Registering Gaia positions on VLBI images}

\author{
M. H. Xu\inst{1}\orcidlink{0000-0001-9602-9489}, 
Y. Y. Kovalev\inst{2}\orcidlink{0000-0001-9303-3263}
}
\authorrunning{Xu \& Kovalev}

\institute{GFZ Helmholtz Centre for Geosciences, 14473 Potsdam, Telegrafenberg, Germany
\email{[minghui.xu@gfz.de]}
\and
Max Planck Institute for Radio Astronomy, Auf dem Hügel 69, D–53121 Bonn, Germany}

\date{Received [date]; accepted [date]}

\abstract
{Because of the complex and time-variable radio structure of active galactic nuclei (AGNs), absolute astrometric positions cannot generally be registered unambiguously with specific features in VLBI images.
The combination of VLBI and Gaia astrometry provides a powerful way to investigate the relationship between radio and optical emission from compact extragalactic sources.
}
{We aim to register Gaia optical centroids on VLBI images with about 0.1~mas accuracy and to locate the optical emission relative to the radio core in the quasar pair B\,1038+528 and B\,1038+529 (known as 1038+528A,\,B), in which the opacity-driven core shift was first discovered.}
{We combine the accurately measured relative position of the pair from phase-referenced VLBI at four epochs between 1981 and 1995, their Gaia optical positions, and the nearly orthogonal radio-jet directions. Assuming that each Gaia--core displacement lies along the local jet axis (that is physically motivated and observationally supported), we register the Gaia centroids relative to the radio cores with about 0.1\,mas accuracy without using absolute VLBI positions.}
{The Gaia centroids lie upstream of the 8.4-GHz radio cores, by 293\,$\pm$\,93\,$\mu$as (3.2$\sigma$) in B\,1038+529 and 169\,$\pm$\,96\,$\mu$as (1.8$\sigma$) in B\,1038+528. The latter agrees with the $\sim$260\,$\mu$as 8.4-GHz core shift implied by the 2.3--8.4 GHz shift measured for this quasar four decades ago. The absolute astrometric position of B\,1038+528 with VLBI lies about 0.6~mas downstream of the core, and its geodetic position time series drifts along the jet by more than 1\,mas, showing that the 0.8--1.6 mas Gaia--VLBI offsets of the pair are dominated by source structure rather than by core shift.}
{Gaia positions can be registered on VLBI images with an accuracy set by the astrometry uncertainties. If the Gaia centroids mark the jet base, the measured separations are absolute 8.4-GHz core shifts, corresponding to projected distances of about 1.2~pc and about 2.5~pc from the jet base; a contribution from an optical jet emission would make them lower limits. The method is applicable to AGNs with Gaia detection and phase-referenced VLBI astrometry and will improve with forthcoming Gaia data releases.}

\keywords{
astrometry -- 
techniques: interferometric -- 
galaxies: jets --
radio continuum: galaxies --
quasars: individual: B1038+528, B1038+529
}

\begin{document}
\maketitle

\section{Introduction}

Active galactic nuclei (AGNs) are powered by accretion onto a supermassive black hole (SMBH). Their continuum emission is a superposition of radiation from several physically distinct regions around this central engine, and each region tends to dominate a different part of the electromagnetic spectrum \citep[e.g.,][]{1995PASP..107..803U,2015ARA&A..53..365N,2017A&ARv..25....2P,2019ARA&A..57..467B}. The optical and ultraviolet continuum of most AGNs is commonly attributed to thermal emission from the accretion disk, extending over some tens to a few hundred gravitational radii from the SMBH \citep{1973A&A....24..337S}. The hard X-ray emission, on the other hand, is usually associated with Comptonization of disk photons in a hot corona located even closer to the black hole \citep{1991ApJ...380L..51H,1993ApJ...413..507H}.
A fraction of AGNs host a powerful relativistic jet. In these objects the jet itself can outshine the disk across a much wider range of the spectrum, from radio through optical, and in blazars even up to X-ray and gamma-ray energies via synchrotron and inverse-Compton processes \citep[e.g.,][]{1998MNRAS.299..433F,2016ARA&A..54..725M,2017A&ARv..25....4B}. Synchrotron radiation from a conical, relativistic jet is self-absorbed so the apparent location of the core, defined as the surface of unity optical depth, recedes towards the SMBH as the observing frequency increases \citep{1979ApJ...232...34B}. As a result, the radio core imaged by VLBI, the optical/ultraviolet continuum from the disk or jet, and high-energy emission from the jet do not, in general, coincide on the sky. Their relative locations encode information about the structure of the innermost, unresolved regions of the AGN \citep[e.g.,][]{2016Galax...4...37M,2019ARA&A..57..467B}.
A direct, purely astrometric handle on this multi-zone picture is to compare the precise sky positions measured for an AGN at different wavelengths. Each astrometric position locates the centroid of the emission dominating that wave-band, although relating that centroid to a particular physical feature is generally non-trivial due to different structure resolutions of various astrometric techniques. 

One way of addressing this challenge is to compare the astrometric positions of AGNs across different wave-bands and localize these positions on the radio images of parsec-scale jets measured by VLBI. This comparison of astrometric positions across different wave-bands has recently become possible for thousands of AGNs thanks to the absolute astrometry done with VLBI at radio and with the Gaia mission at optical wave-bands. Geodetic/astrometric VLBI over the last five decades has established a celestial reference frame consisting of thousands of AGNs with an accuracy at the sub-mas level \citep[e.g., ICRFs and RFC,][]{2020A&A...644A.159C,2025ApJS..276...38P}. 
In the following the ICRF3 astrometric positions, adopted by the IAU, are used for studying AGN position offsets between radio and optical.
We note that the method to locate optical centroids on VLBI images in this study does not rely on absolute VLBI positions.

The European Space Agency Gaia mission, through its Early Data Release 3 \citep[EDR3;][]{2023A&A...674A...1G}, provides astrometric positions of 1.6 million quasar-like objects (mostly AGNs) with a comparable accuracy to the ICRF \citep{2022A&A...667A.148G}. A comparison between VLBI and Gaia astrometric positions immediately allows the AGNs with large position differences to be identified, about 10\% of the common objects between VLBI and Gaia showing statistically significant offsets \citep[see, e.g.,][]{2017MNRAS.467L..71P,2019MNRAS.482.3023P,2019ApJ...871..143P,2019ApJ...873..132M,2022ApJ...939L..32S}. More interestingly, for the majority of these objects with significant offsets the position offsets are aligned to the respective radio jet directions, first reported by \citet{2017A&A...598L...1K} and confirmed by, for example, \citet{2019MNRAS.482.3023P,2021A&A...647A.189X}. Optical jet emission is commonly suggested as the dominant contributor to the downstream offsets \citep[see][]{2019ApJ...871..143P}, while optical AGN flares affect Gaia centroid motion \citep{2026ApJ..1006L..27P}. However, the remaining majority of the objects have much smaller position offsets at the sub-mas level and it is challenging to investigate their VLBI--Gaia position offsets and localize the optical positions with respect to the radio structure. 
On the other hand, such sources may be even more interesting, because their Gaia centroids are less likely to be dominated by extended optical jets and may therefore trace regions closer to the central engine. Their optical emission can come either from the accretion disk, which has been shown to drive, at least in part, the offsets observed upstream of the radio jet \citep{2019ApJ...871..143P}, or from the jet base, since electrons radiating in the optical cool much faster than those radiating in the radio and the optical jet is therefore expected to be short unless particles are reaccelerated downstream \citep{2009ApJ...696.1142M,2026A&A...711A.227F}.

Localization of Gaia positions can be possible at the mas level, which already allows for interesting studies. For instance, by exploring this possibility for the AGNs with the Gaia--VLBI position offsets larger than 1.0\,mas, \citet{2025MNRAS.543..479P} identified 35 AGNs for which the VLBI position corresponds to a jet component and the Gaia position is associated with the core, which is weaker than the jet, and \citet{2025A&A...704A..93A} used the Gaia position as an indicator of the radio core to search for compact symmetric objects. \citet{2024A&A...684A.202L} attempted to identify the optical emission based on the 15\,GHz images for hundreds of AGNs. However, the localization of Gaia positions with an accuracy of $\sim0.1$\,mas has not been achieved so far.

Two main effects contribute to the radio and optical position offsets at the sub-mas level, which prevent such precise localization. The first effect is the time- and frequency-dependent angular structure of AGNs. \citet{2025AJ....169..173X} have argued that the contribution of source structure to the Gaia--VLBI position offsets is about 0.5\,mas for the geodetic/astrometric VLBI systems, which observe at frequencies from 2 to 11 GHz. The second effect is core shift: the frequency-dependent displacement of the core position along the jet due to the opacity effect in synchrotron radiation \citep{1998A&A...330...79L}. It arises because relativistic plasma can be opaque to its own synchrotron emission. The magnitude of core shifts has been measured for a fairly large sample of AGNs; they are on the order of 0.4\,mas between 2.3 and 8.4 GHz and 0.1\,mas between 8.4 and 15 GHz \citep[e.g.,][]{2008A&A...483..759K,2011A&A...532A..38S,2012A&A...545A.113P,2019MNRAS.485.1822P}. In fact, these two effects are highly coupled. \citet{porcas2009} argued that, for a point-like AGN (by assumption), due to core shift the absolute astrometric position determined from geodetic/astrometric VLBI marks the location of the jet base rather than the apparent location of the core at the observing frequency. In this paper, we use the term ``jet base'' to refer to the true jet apex, which effectively is the core position extrapolated to infinite frequency; the jet base itself is not directly visible because of opacity. Other factors can also lead to the Gaia and VLBI position offsets in different ways. The existence of optical jets contributes to the offsets in the jet directions, whereas the host-galaxy contamination shifts the offsets with no preferred directions. In general, their probabilities need to be evaluated case by case.

\begin{figure}[b]
\centering
\includegraphics[width=1.0\linewidth]{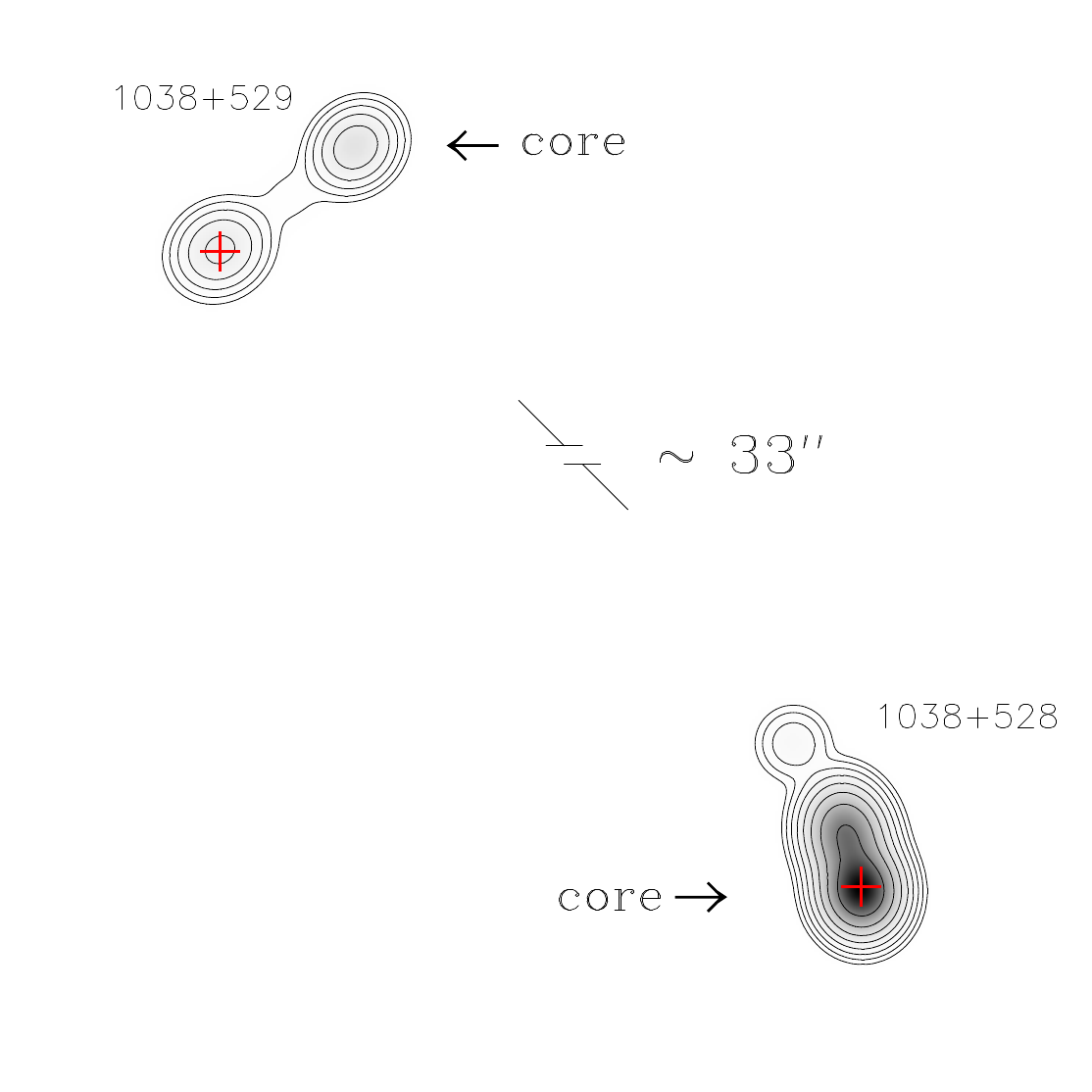}
\caption{\label{fig:pair}Source pair of quasars B\,1038+528 and B\,1038+529 that are $\sim$33 arcseconds apart on the sky. Their images were originally derived from 8.4\,GHz observations in November 1995; they are created based on the Gaussian models listed in Table~1 of \citet{2000A&A...355..552R} with a circular restoring beam of 0.5\,mas. The contours present the relative strength between the two quasars, starting at 0.4\% of the peak intensity and increase by a factor of 2. The reference points used in the relative astrometry are indicated by red pluses.}
\end{figure}

In this study, we present a method to localize the absolute astrometric positions from both VLBI and Gaia on radio images. To achieve the desired high accuracy this method makes use of relative astrometric positions measured with VLBI phase referencing technique, which derives a relative position between identifiable image features, with the structure absorbed in the calibrator model during data calibration. While this method can be applied to a sample of AGNs with Gaia detection and phase referencing observations, we first test it on a pair of quasars, B\,1038+528 and B\,1038+529, which are separated by only 33$''$ on the sky, and report their results. By discussing the implication of the results we argue that this method can be a novel approach to localize the optical emission of AGNs and measure core shifts.
Throughout this paper we adopt a flat $\Lambda$CDM cosmology with 
$H_0 = 67.7$\,km\,s$^{-1}$\,Mpc$^{-1}$, $\Omega_\mathrm{m} = 0.31$, and 
$\Omega_\Lambda = 0.69$ \citep{2020A&A...641A...6P}, for which an angular 
separation of 1\,mas corresponds to a projected linear distance of 7.26\,pc 
at the redshift of B\,1038+528 ($z = 0.678$) and 8.41\,pc at that of 
B\,1038+529 ($z = 2.296$).

\section{Quasar pair B\,1038+528 and B\,1038+529}

The quasar pair B\,1038$+$528 and B\,1038$+$529 (originally known as 1038+528A,\,B), discovered by \citet{1980ApJ...235L..57O}, have optical $G$ magnitudes of 17.1 and 18.2 \citep{2022A&A...667A.148G} with redshifts of 0.678 \citep{1980ApJ...235L..57O,2018A&A...613A..51P} and 2.296 \citep{1980ApJ...235L..57O,2015ApJS..219...12A}, respectively. The angular separation between them on the sky is only $\sim$33$^{\prime\prime}$, as illustrated in \autoref{fig:pair}. They are thus ideal candidates for high-accuracy phase-referencing observations, even successfully carried out with Earth-Space VLBI \citep{2000AdSpR..26..673P}. Upon the proposal by \citet{1980ApJ...235L..57O} to use this pair for testing cosmological versus local quasar hypotheses via relative proper motion, over the time period 1981--1995 the two quasars were observed using phase referencing for multiple epochs \citep{1983AJ.....88.1133M,1994AJ....108..368M, 1997A&A...325..383R,2000A&A...355..552R}. At 2.3\,and 8.4\,GHz B\,1038+528 has a total flux density of about 600\,mJy whereas B\,1038+529 is about an order of magnitude weaker. They have prominent jet components detected at scales up to a few mas, with the jet position angles of 20.1$^{\circ}$\,$\pm$\,3.0$^\circ$ for B\,1038+528 and 126.9$^{\circ}$\,$\pm$\,0.2$^\circ$ for B\,1038+529. Through these phase referencing observations a frequency-dependent displacement of the peak brightness location was identified and confirmed \citep{1983AJ.....88.1133M,1994AJ....108..368M, 1997A&A...325..383R,2000A&A...355..552R}, which was considered to be the first observational evidence of the core shift effect \citep{1984ApJ...276...56M}. 

\begin{figure}
\centering
\includegraphics[width=1.0\linewidth]{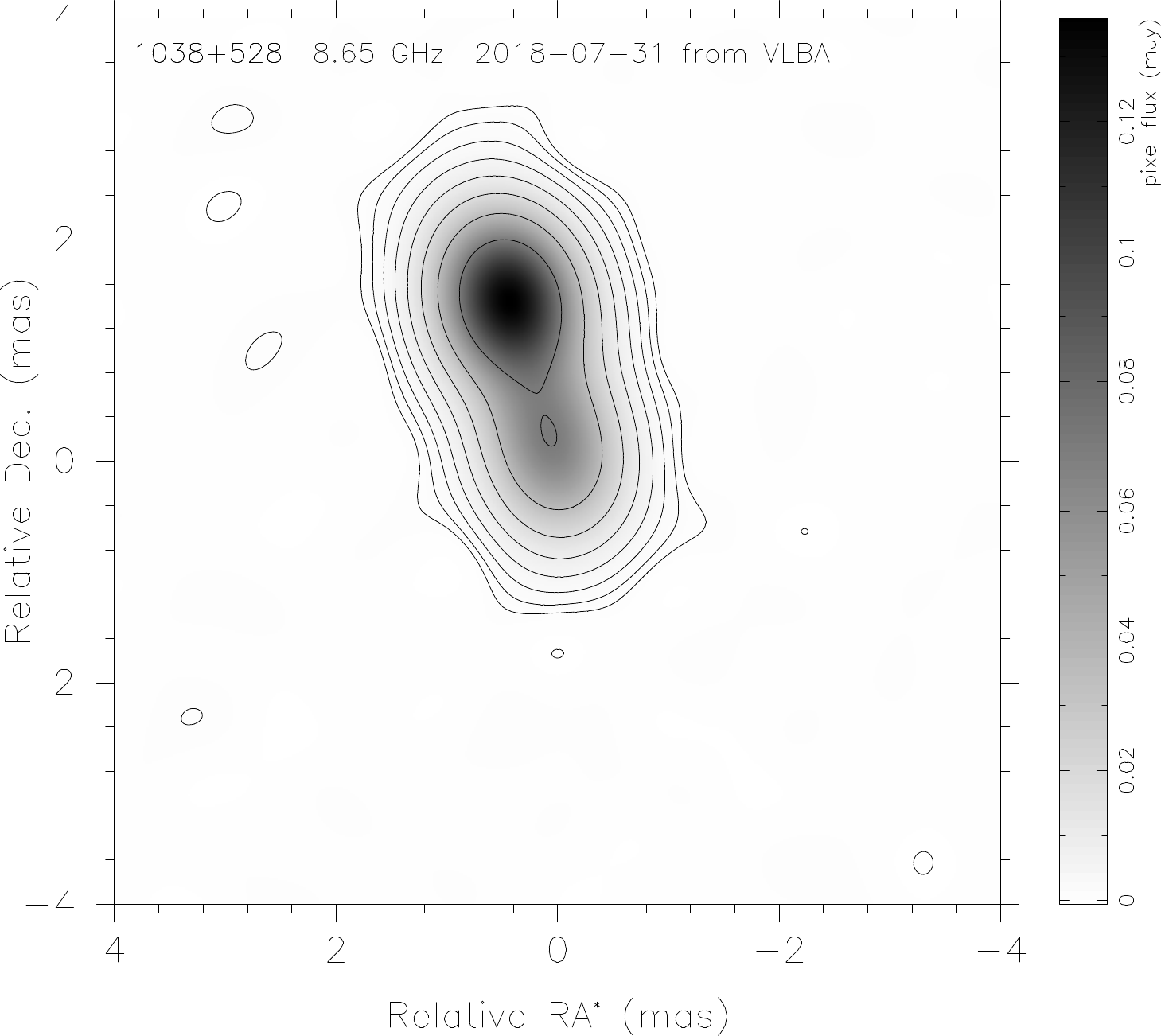}
\caption{\label{fig:1038+528_2018} 
VLBA CLEAN image of quasar B\,1038+528 at 8.65~GHz, observed on 2018 July 31. The contours start at 0.4\% of the peak intensity 589~mJy/beam and increase by a factor of 2. The restoring beam has an FWHM of $1.366\,\times\,0.968\,\mathrm{mas}$, with a major-axis position angle of $10.31^\circ$. The image is sampled with a pixel $15\times15\,\mu$as while plotting the gray scales.}
\end{figure}

The Gaia--VLBI position offsets for both quasars are of the same order of magnitude as the angular distances between the cores and jets. The jet components of quasar B\,1038+528 have evolved since 1996 to be stronger than the core and also move away from the core, as shown in its image at epoch 2018\,July\,31 in \autoref{fig:1038+528_2018}. For the quasar B\,1038+529, the brightness peak corresponds to the jet component at 2.3\,GHz and to the core at 8.4\,GHz. The jet component was selected as the reference in relative astrometry because it provided a consistent, optically thin reference at both bands. Although the Gaia positions with respect to the VLBI positions are in the opposite directions to their radio jets, it is impossible to infer whether they are on the jet side or the other side with respect to the cores based merely on the absolute astrometric positions. The relative astrometric position sheds light on this question.

\section{Astrometric positions}\label{sec:astrometry}

The relative position of the quasar pair was measured at four epochs with an accuracy of about 30\,$\mu$as per coordinate (see \autoref{tab:phase}). For these reported relative positions, the reference points used were the core for B\,1038+528 but a jet component for B\,1038+529~--- that was much brighter than the core at S band. It was concluded that the motion of the jet component (the reference point) of B\,1038+529 has a speed of 13--17\,$\mu$as\,yr$^{-1}$ and that of the core was less than 5\,$\mu$as\,yr$^{-1}$ \citep{2000A&A...355..552R}. Therefore, we calculate the relative position referring to the cores of both quasars by subtracting the position offset between the core and the jet of B\,1038+529 determined from the image. This calculation is done epoch by epoch based on the jet-to-core separation reported in Table 1 in both \citet{1997A&A...325..383R} and \citet{2000A&A...355..552R} and the standard deviation of the resultant core-to-core positions at the four epochs agrees with the position uncertainties from individual epochs when accounting for the potential core motion of B\,1038+529. 

\begin{table}[tb]
\caption{Relative positions of the jet of B\,1038+529 with respect to the core of B\,1038+528 at 8.4\,GHz determined from VLBI phase reference measurements.}
\label{tab:phase}
\centering
\begin{tabular}{ccccccc}
\hline\noalign{\smallskip}
Epoch& $\Delta\alpha$ & $\sigma$($\Delta\alpha$) & $\Delta\delta$ & $\sigma$($\Delta\delta$)& Ref.\\
&&($\mu$as)&&($\mu$as)&\\
\noalign{\smallskip}\hline\noalign{\smallskip}
1981.2&2$^s$.1160427& 25 &27$^{\prime\prime}$.376574 & 36 & (1) \\
1983.4&2$^s$.1160439& 24 &27$^{\prime\prime}$.376510 & 34 & (2) \\
1990.5&2$^s$.1160568& 28 &27$^{\prime\prime}$.376464 & 36 & (3) \\
1995.9&2$^s$.1160588& 27 &27$^{\prime\prime}$.376325 & 25 & (4) \\
\noalign{\smallskip}\hline
\end{tabular}
\begin{tablenotes}
\item \textit{Note}. References: (1)\,\cite{1983AJ.....88.1133M}; (2)\,\cite{1994AJ....108..368M}; (3)\,\cite{1997A&A...325..383R}; (4)\,\cite{2000A&A...355..552R}.
 \end{tablenotes}
\end{table}

\begin{table*}[tb]
\caption{Absolute positions of the quasars B\,1038+528 and B\,1038+529 from Gaia \citep{2022A&A...667A.148G} and geodetic/astrometric VLBI \citep{2020A&A...644A.159C}.}
\label{tab:absolute}
\centering
\begin{tabular}{ccccrccrrc}
\hline\noalign{\smallskip}
\multicolumn{2}{c}{Source Name/ID} && $\alpha$ & $\sigma_\alpha\cos\delta$ && $\delta$ & $\sigma_\delta$ & C$_{\alpha\delta}$ & RUWE\\
&&& &($\mu$as)&&&($\mu$as) &\\
\noalign{\smallskip}\hline\noalign{\smallskip}
Gaia & DR3 849141633152788096 &&10$^h$ 41$^m$ 46$^{s}$.7816366 & 37 & &  52$^\circ$ 33$^\prime$ 28$^{\prime\prime}$.231026 &  50 & $-$0.19 & 0.98\\
VLBI & B\,1038+528 &&10$^h$ 41$^m$ 46$^{s}$.7816540 & 34 & & 52$^\circ$ 33$^\prime$ 28$^{\prime\prime}$.231796 &  36 & +0.03 & --\\
\noalign{\smallskip}\hline
Gaia & DR3 849141633152787840 & &10$^h$ 41$^m$ 48$^{s}$.8975105 & 71 & &52$^\circ$ 33$^\prime$ 55$^{\prime\prime}$.608906 &  89 & $-$0.19 & 1.00\\
VLBI & B\,1038+529   &&10$^h$ 41$^m$ 48$^{s}$.8976707 & 396 & & 52$^\circ$ 33$^\prime$ 55$^{\prime\prime}$.608201 &  355 & $-$0.45 & -- \\
\noalign{\smallskip}\hline
\end{tabular}
\end{table*}

For subsequent analysis, we use these four-epoch measurements to obtain the averaged core-to-core position, which is 
\[
\begin{aligned}
\quad & \Delta\alpha = 2^{\mathrm s}.1158933\,{\pm}\,13\,\mu{\mathrm as}, & \Delta\delta &= 27''.377545\,{\pm}\,27\,\mu{\mathrm {as}}. \\
\end{aligned}
\]
The absolute astrometric positions from Gaia and VLBI are listed in \autoref{tab:absolute}. Gaia provides the absolute astrometric positions with uncertainties of tens of $\mu$as, and any potential errors from frame tie (between radio and optical) and Gaia large-scale systematics cancel at a 33$''$ separation. As reported in the third realization of the ICRF \citep[ICRF3,][]{2020A&A...644A.159C}, geodetic/astrometric VLBI measures the astrometric position of quasar B\,1038+528 with an uncertainty comparable to that of Gaia, but for the weaker quasar B\,1038+529 the astrometric uncertainty is one order of magnitude higher. The position offsets between VLBI and Gaia are 0.79\,$\pm$0.07\,mas and 1.63$\pm$0.55\,mas for B\,1038+528 and B\,1038+529 respectively. The position angles of the Gaia--VLBI position vectors are 11.7$\pm$2.3$^\circ$ for B\,1038+528 and 115.8$\pm$12.2$^\circ$ for B\,1038+529, which are aligned to their parsec-scale jet directions. 

Previous studies have demonstrated that the absolute VLBI position estimates of an AGN vary along its jet direction \citep[see, e.g.,][]{2024ApJ...977L..14M,2025AJ....169..173X}. For B\,1038+528, which has a long history of geodetic VLBI observations, this behavior can be directly seen from the position time series presented in \autoref{structure_effect} (see \autoref{fig:1038+528_geodesy}). Therefore the observed alignment of the Gaia--VLBI offset vectors with the jet directions for this quasar pair provides empirical evidence that the Gaia position relative to the core also lies in the jet direction, either downstream or upstream. This is further supported by their cosmological redshifts. At such distances, the optical emission from the disk should sit below 1\,$\mu$as and the compact optical synchrotron emission should be within a few tens of $\mu$as of the apex, with the major possible factor being the optical jet. We use this observational constraint and the physically motivated argument in subsequent analysis. 

\section{Registering the absolute astrometric positions on the images}\label{sec:register}

We exploit two observed properties: (1) the jet axes of the two quasars are nearly perpendicular and (2) their Gaia–VLBI offset vectors are aligned to their respective jet directions. The near-orthogonality makes the system well-conditioned (for parallel jets it would be singular), while the second constrains each Gaia-to-radio-core displacement to lie along the corresponding jet axis. The signed angular separation of the Gaia position with respect to the radio core, denoted by $r_{\mathrm a}$ for B\,1038+528 and $r_{\mathrm b}$ for B\,1038+529, can be obtained from the following equations
\begin{equation}\label{eq1}
\begin{gathered}
    r_{\mathrm a} = \frac{ (\Delta\alpha_{\mathrm{ab}}^{\mathrm{Gaia}} - \Delta\alpha_{\mathrm{ab}}^{\mathrm{phase}}) \cos\theta_{\mathrm b} - (\Delta\delta_{\mathrm{ab}}^{\mathrm{Gaia}} - \Delta\delta_{\mathrm{ab}}^{\mathrm{phase}}) \sin\theta_{\mathrm b}}{\sin(\theta_{\mathrm b} - \theta_{\mathrm a})},   \\
        r_{\mathrm b} = \frac{ (\Delta\alpha_{\mathrm{ab}}^{\mathrm{Gaia}} - \Delta\alpha_{\mathrm{ab}}^{\mathrm{phase}}) \cos\theta_{\mathrm a} - (\Delta\delta_{\mathrm{ab}}^{\mathrm{Gaia}} - \Delta\delta_{\mathrm{ab}}^{\mathrm{phase}}) \sin\theta_{\mathrm a}}{\sin(\theta_{\mathrm b} - \theta_{\mathrm a})},  
        \end{gathered}
\end{equation}
where ($\Delta\alpha_{\mathrm{ab}}^{\mathrm{Gaia}}$, $\Delta\delta_{\mathrm{ab}}^{\mathrm{Gaia}}$) is the relative Gaia position of B\,1038+529 with respect to B\,1038+528, ($\Delta\alpha_{\mathrm{ab}}^{\mathrm{phase}}$, $\Delta\delta_{\mathrm{ab}}^{\mathrm{phase}}$) is the relative position between the two radio cores from phase referencing observations ($\Delta\alpha$ denotes the projected right-ascension difference in the tangent plane), and $\theta_{\mathrm a}$ and $\theta_{\mathrm b}$ are the position angles of the radio jets for B\,1038+528 and B\,1038+529, respectively. In deriving Eqs.~\ref{eq1}, the displacement between each Gaia centroid and its corresponding radio core is taken to be parallel to the local radio-jet axis, as inferred from the observed alignment (see \autoref{sec:astrometry}). The two parameters, $r_{\mathrm a}$ and $r_{\mathrm b}$, are defined as positive when the Gaia position is downstream with respect to the core. The uncertainties of the jet position angles, the full covariance of the Gaia optical positions and that of the radio relative astrometry are used to compute the uncertainties of the two parameters through both analytic error propagation and Monte Carlo simulations.

\begin{figure}[tb]
\centering
\includegraphics[width=1.0\linewidth]{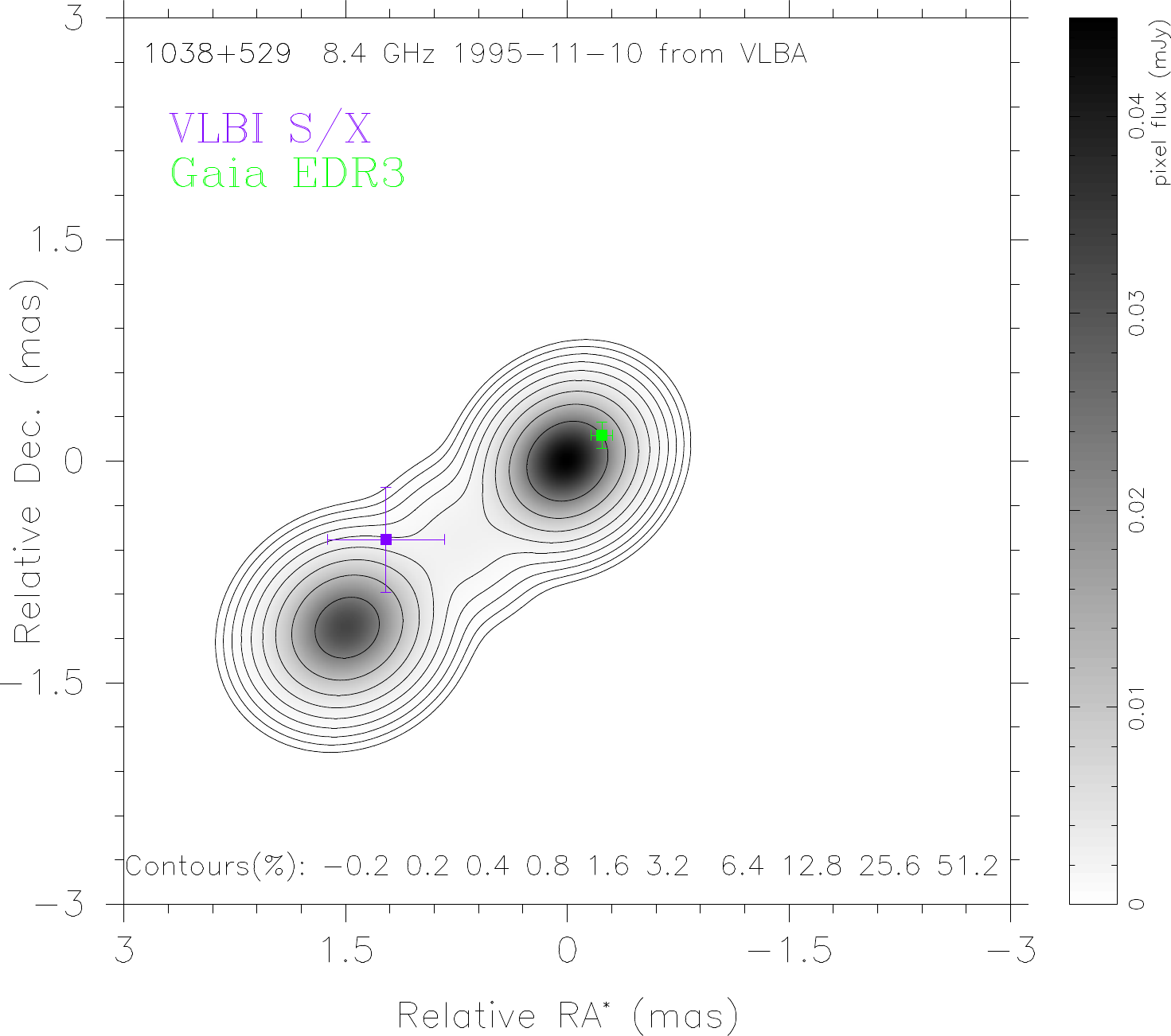}
\includegraphics[width=1.0\linewidth]{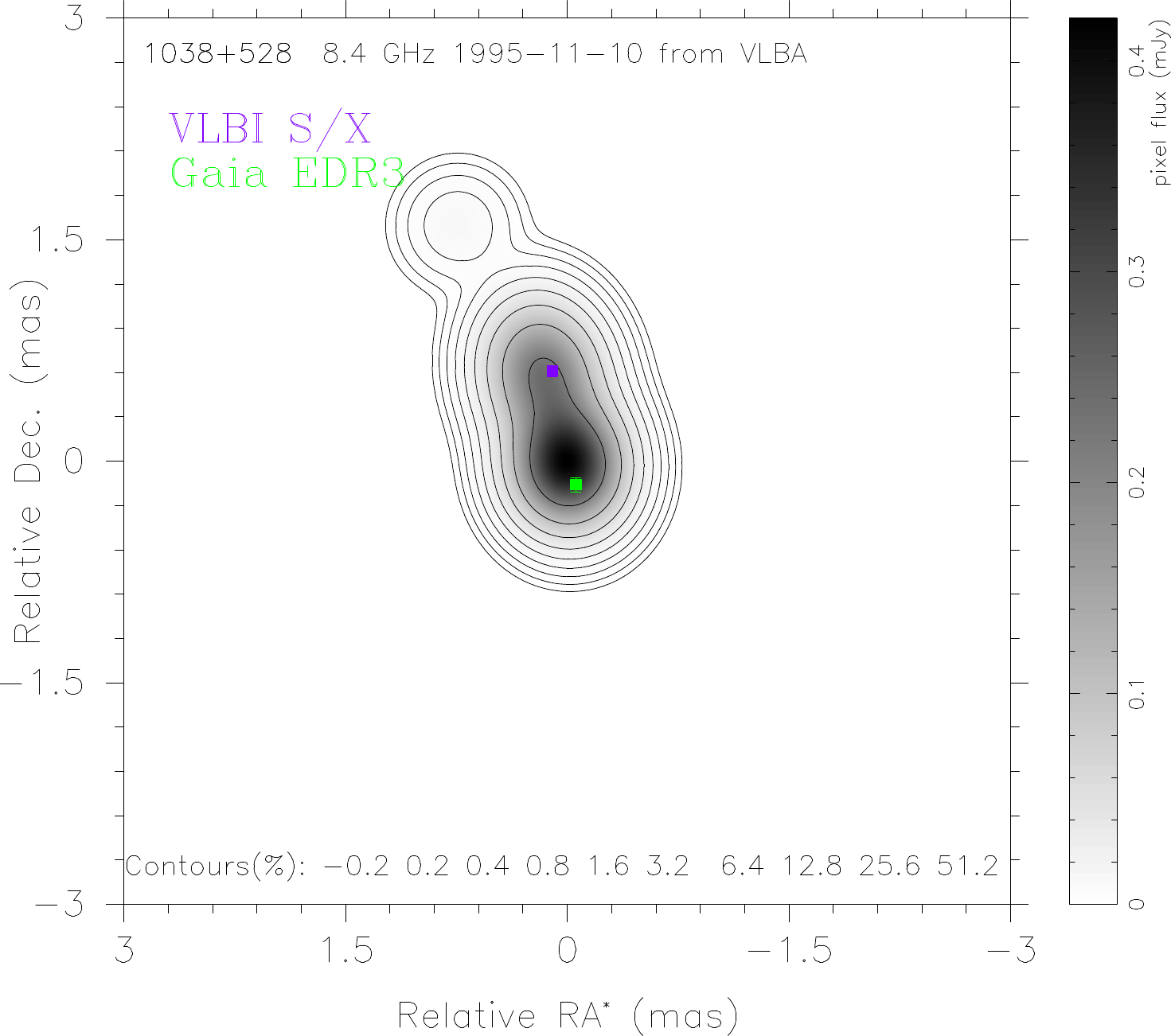}
\caption{\label{fig:opacity}
The absolute astrometric positions from Gaia and VLBI (S/X\,band) are registered on the model images at 8.4\,GHz on 1995 November 10. The origins of the two images correspond to the locations of the radio cores. These images are created from the combination of Gaussian models listed in Table~1 of \citet{2000A&A...355..552R} with a circular restoring beam of 0.5\,mas. These models were obtained from the same dataset of phase referencing observations, from which the relative position between the radio cores was measured by \citet{2000A&A...355..552R} and is used in this study. We assume that the radio cores do not move on the sky.}
\end{figure}

Based on the positions listed in Tables~\ref{tab:phase} and \ref{tab:absolute}, the solution yields the result $r_{\mathrm a} = -169 \pm 96$\,$\mu$as and $r_{\mathrm b} = -293 \pm 93$\,$\mu$as with the uncertainties evaluated through Monte Carlo simulations. According to the definition of the parameters in Eqs.~\ref{eq1} the negative signs of the estimates indicate that the Gaia position with respect to the radio core is in the opposite direction to that of the radio jet for both quasars, i.e.\ upstream. Through Monte Carlo simulations of 200,000 runs the uncertainties of the parameter estimates caused by the errors of the jet position angles are a few $\mu$as; even for jet angle errors with a magnitude of 5\,$^\circ$ the impact is $\sim$20\,$\mu$as. The simulations demonstrate that the Gaia position uncertainties in the current Gaia data release dominate ($>$90$\%$) the uncertainties of the results. The inferred parameters have a low correlation coefficient of $-0.1$, thanks to the nearly perpendicular jet position angles between the two quasars. This low correlation can also be achieved when three or more AGNs are observed using relative astrometry.
The ICRF3 position lies downstream of the 8.4-GHz core by 0.62\,mas at position angle of 9$^{\circ}$ for B\,1038+528 and by 1.34\,mas at position angle of 113$^{\circ}$ for B\,1038+529. The relative geometry between the radio core and the absolute astrometric positions from Gaia and geodetic/astrometric VLBI observations at S/X band ($\sim2\,/\,8$~GHz) is shown in \autoref{fig:opacity}.

\section{Discussion}

\subsection{Apparent motions of the radio cores}
The reference epoch of the Gaia EDR3 astrometry is J2016.0 whereas the relative radio core-to-core position of the quasar pair was measured during the time period 1981--1995. A downstream motion of 3.8 $\pm$ 0.3\,$\mu$as\,yr$^{-1}$ was estimated for the core of quasar B\,1038+529 although a statistically insignificant motion of 5.5\,$\pm$\,3.6\,$\mu$as\,yr$^{-1}$ was estimated for B\,1038+528 \citep{2000A&A...355..552R}. We note that the Gaia EDR3 cannot independently verify these apparent motions at the $\mu$as\,yr$^{-1}$ level. The apparent motions of the cores might change the Gaia-to-radio-core displacement at the epoch J2016.0. However, \citet{2000A&A...355..552R} attributed the measured apparent motions to time-variable structure at spatial scales below the resolution of the VLBI imaging data rather than the secular motion of the quasar itself. The unresolved inner jet typically moves further out, becomes resolved, and thus does not cause a long term motion of the core; it is not justified to propagate this apparent motion to the Gaia reference epoch. We therefore assume that any long-term physical displacement of the core is negligible compared with the astrometric uncertainty, and the references for locating the Gaia positions are in fact the mean core positions over the four epochs in 1981--1995. Propagating the reported apparent motions to J2016.0 would have increased the upstream Gaia-to-radio-core separations by 100--130\,$\mu$as, thereby increasing rather than decreasing their statistical significance. However, caveat to this interpretation is that the 8.4-GHz core may be blended due to the inner jet as noted.

\subsection{Opacity-driven core shift}

The Gaia positions lie upstream of the 8.4-GHz radio cores for both quasars, i.e. in the directions opposite to their respective radio jets. This geometry is consistent with the Gaia centroid being dominated by emission from the vicinity of the central engine rather than by optical jet emission downstream of the radio core. If this interpretation is correct, the Gaia position can serve as an approximation to the jet base, i.e., the radio-core position extrapolated to infinite frequency. The measured Gaia-to-radio-core separation can then be interpreted as the opacity-driven displacement of the 8.4-GHz core.

The angular separation between the Gaia position and the radio core that we measured can be interpreted as a core shift of 293\,$\pm\,93$\,$\mu$as for B\,1038+529 at 8.4\,GHz. For quasar B\,1038+528, a tentative displacement is measured to be 169\,$\pm$\,96\,$\mu$as. Based on the core shift of 0.7\,mas between 2.3\,GHz and 8.4\,GHz for quasar B\,1038+528 as originally reported using relative astrometry alone \citep{1983AJ.....88.1133M,1984ApJ...276...56M,1994AJ....108..368M}, the 8.4\,GHz core should have a shift of $\sim$263\,$\mu$as assuming that the core shift depends on frequency as $r\propto\nu^{-1}$ \citep{1998A&A...330...79L,2011A&A...532A..38S}. This is consistent with our result of B\,1038+528 within 1$\sigma$. 
With the adopted cosmology, these separations correspond to projected 
distances of the 8.4-GHz cores from the jet base of $1.2\pm0.7$\,pc for 
B\,1038+528 and $2.5\pm0.8$\,pc for B\,1038+529.
The deprojected distances are larger by a factor of $1/\sin\theta$, 
where $\theta$ is the jet viewing angle; this factor is typically $\sim10$ for blazars.
We conclude that the method provides a new approach for measuring core shifts in samples of Gaia-detected AGNs for which phase referencing observations are available.

\begin{figure}
\centering
\includegraphics[width=1.0\linewidth]{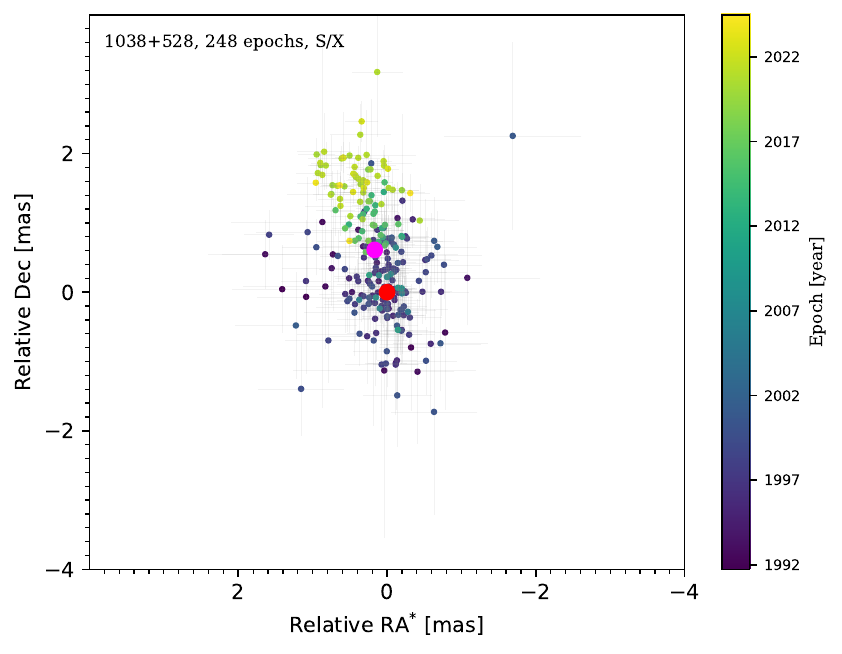}
\caption{\label{fig:1038+528_geodesy} Sky distribution of the residuals of the astrometric position estimates of B\,1038+528 from geodetic/astrometric VLBI observations since 1980. These position residuals refer to the absolute position of the core (indicated by red dot) as determined from \autoref{sec:register}, with respect to which the ICRF3 position is also localized as shown by magenta dot.}
\end{figure}

\subsection{Impact of source structure}\label{structure_effect}

The quasar B\,1038+529 has a jet component as bright as its core. During the period from 1981 to 1995, B\,1038+528 exhibited source structure that is typical in geodetic/astrometric VLBI observations -- a bright, compact core accompanied by extended, weaker jet components. An interesting question is why the absolute VLBI position of B\,1038+528 is located closer to a jet component that was always much weaker than the core during the decade of the phase referencing observations. To demonstrate the trajectory of the astrometric position changes, we performed a global solution using geodetic/astrometric VLBI observations at S/X bands since 1980 \citep{2017JGeod..91..711N} and solved for the position time series of the quasar B\,1038+528 while estimating the positions of all other sources as global parameters \citep[see][]{2025AJ....169..173X}. \autoref{fig:1038+528_geodesy} shows the sky distribution of the position estimates at 248 epochs. The astrometric position estimates of B\,1038+528 move from the southwest to the northeast. This apparent motion is consistent with its structure evolution, i.e., the jet components brighten to be stronger than the core and move outward (refer to its image at epoch 2018 July 31 in \autoref{fig:1038+528_2018}). The results address the question: the geodetic data volume is dominated by post-2000 sessions, when the jet components dominated. 

The sky distribution of the position estimates shows an elongation excess (the ratio between the major axis and the minor axis of the sky distribution) of 4.7 $\pm$ 0.5 towards the major-axis direction of 18.5$^\circ$ $\pm$ 1.7$^\circ$, with the uncertainties evaluated through Monte Carlo bootstrapping \citep{2024ApJ...977L..14M}; the elongation direction in the astrometric position variations is consistent with the jet direction. The proper motion derived from the position time series is 13.8\,$\pm$\,2.1\,$\mu$as/yr in right ascension and 61.4\,$\pm$\,1.6\,$\mu$as/yr in declination. Structure evolution of B\,1038+528 changes the astrometric positions by more than 1.0\,mas over several decades.

For this quasar pair, the 0.8--1.6~mas Gaia–VLBI offsets appear to result from the combined effects of source structure and core shift. The core shift contributes up to 0.3\,mas, whereas the effects of source structure contribute approximately 0.5--1.0\,mas. The absolute astrometric position as reported in the ICRF3 corresponds to the ``mean'' location of these estimates over the full observing history, whereas the individual astrometric position estimates evolve systematically with time because of the motion and changing brightness of the dominant jet.

As argued by \citet{porcas2009}, VLBI astrometric positions derived from group delays, such as those in the ICRF3, may correspond to the jet base and therefore lie closer to the Gaia positions than to the radio core positions. In this case, core shift may contribute less than the 0.3\,mas to the Gaia--VLBI position offsets, as revealed by the phase delay positions measured with phase referencing, implying that the contribution of source structure can even increase the magnitude. Due to the coupling of core shift and source structure, however, a quantitative evaluation to verify this implication requires modeling of source structure of AGNs. Modeling source structure in astrometric/geodetic VLBI data analysis when deriving the absolute positions is one of the objectives in our ongoing research work \citep{2021JGRB..12621238X,2022A&A...663A..83X}; after this effort, the effect in \citet{porcas2009} can be investigated and will be reported.

\section{Summary}

We have developed a method using the relative astrometric positions from VLBI phase referencing observations to locate the optical positions of AGNs from Gaia on their radio images as well as the absolute positions derived from VLBI. In addition to the high accuracy of the relative astrometry, this approach benefits from the fact that the effects of source structure were already accounted for during visibility data calibration, which provides a clear identification of the reference points of the relative astrometry on the radio images. This geometric method can localize the optical-emission centroid on a VLBI image with a precision comparable to the underlying astrometric uncertainties. It thus enables the VLBI images to be registered with absolute positions and the emission centroids measured at different wavebands (radio and optical) to be placed on these VLBI images with high accuracy. It will provide further insight into the origin of Gaia--VLBI position offsets by applying this method to a large sample of AGNs that have phase-referencing observations.

We tested this method on the quasar pair B\,1038+528 and B\,1038+529, for which more than a decade of phase referencing observations are available. This is the first study to locate the optical-emission centroids of AGNs relative to their radio structure with an uncertainty of 0.1\,mas. This high-accuracy localization opens up important astrophysical applications. For this quasar pair, the optical centroids are located upstream of the radio cores. This finding is consistent with optical emission originating near the central engine. When optical-jet contamination is negligible, its separation from the radio core can therefore constrain the opacity-driven core shift; if not, the separation that we measure provides a lower limit. The current methods widely used for measuring core shifts primarily determine relative shifts between radio frequencies \citep[e.g.,][]{2008A&A...483..759K,2011A&A...532A..38S,2015AJ....150..202R}, whereas the approach developed here registers the Gaia optical position relative to a specific radio core.

We infer an 8.4-GHz core displacement of \(293\pm93\ \mu\mathrm{as}\) (3.2$\sigma$) due to opacity for B\,1038+529 under the assumption that the Gaia centroid originates from the vicinity of the central engine. This result also independently confirms the identification of the core for quasar B\,1038+529, which is non-trivial for AGNs with two equally bright and compact features \citep[see][]{2025MNRAS.543..479P}. A tentative displacement of 169\,$\pm\,$96\,$\mu$as (1.8$\sigma$) is measured for the quasar B\,1038+528; this constraint can be improved with future Gaia data releases as the uncertainties of the Gaia EDR3 are the dominant errors in this case. Compared with core shifts the Gaia--VLBI position offsets (based on absolute astrometry) are more strongly affected by source structure and its evolution. The impact on the mean position coordinates can reach 0.5~mas or more, even for a typical AGN appearing with a dominant core and weak jet morphology over multiple years. 

In future work, this approach has the potential to localize the Gaia positions on radio images in more general cases where the relative astrometric positions from phase referencing are available or can be measured, preferably with nearly perpendicular jets or groups of more than two AGNs. Possible astrophysical applications enabled by this localization of the optical emission include measuring core shifts, identifying compact symmetric objects \citep{2024ApJ...961..240K,2025A&A...704A..93A}, and investigating gravitational lensing by localizing the optical centroids of the multiple images precisely \citep{2015aska.confE..84M,xu2022}. 

\section*{Acknowledgments}

We thank Richard Porcas for helpful comments and Petra Benke for discussions.
This research was funded by the European Union (ERC Astrogeodesy project No 101076060 and MuSES project No 101142396). Views and opinions expressed are, however, those of the author(s) only and do not necessarily reflect those of the European Union or the European Research Council. Neither the European Union nor the granting authority can be held responsible for them.
We used in our work the Astrogeo VLBI FITS image database, maintained by Leonid Petrov. 
        
\section*{Data Availability}

The total intensity image of B\,1038+528 at epoch 2018 July 31 that is used in our work is publicly available in the Astrogeo VLBI FITS image database 
\href{https://doi.org/10.25966/kyy8-yp57}{DOI: 10.25966/kyy8-yp57}.

\bibliographystyle{aa}
\bibliography{gaia_crf}

\end{document}